\documentstyle[11pt,newpasp,twoside,psfig]{article}
\begin{document}

\title{Molecular outflow in the composite galaxy NGC~6764}
\author{St\'ephane Leon}
\affil{Instituto de Astrof\'{\i}sica de Andaluc\'{\i}a, Camino Bajo de Hu\'etor, 24
Apdo 3004, 18080 Granada, Spain}
\author{Andreas Eckart}
\affil{I. Physikalisches Institut, University of Cologne, Cologne, Germany}
\author{Seppo Laine}
\affil{SIRTF Science Center, California Institute of Technology, 220-6,
Pasadena, CA 91125 }
\author{Eva Schinnerer}
\affil{Department of Astronomy, California Institute of Technology, Pasadena, CA 91125-2400, USA }

\setcounter{page}{111}
\index{Leon, S.}
\index{Eckart, A.}
\index{Laine, S.}
\index{Schinnerer, E.}

\begin{abstract}
NGC~6764 is a composite galaxy exhibiting central activity (LINER) and recent
massive star formation (Wolf-Rayet galaxy). We report new high-resolution interferometric data 
of the  $^{12}$CO(1--0)  and $^{12}$CO(2--1) emission in the central part of the galaxy together 
with high resolution VLA data of the radio continuum emission. The bulk of molecular gas (few 
times $10^8 {\rm \ M_{\odot}}$)  
is  located in the central part. A molecular outflow ($\sim  1.7\times 10^7 {\rm \ M_{\odot}}$)
out-of the plane is associated with the starburst-driven outflow radiocontinuum emission. 
\end{abstract}

\section{Introduction}
NGC~6764 is a barred spiral galaxy classified as a low-ionization nuclear emission-line
region (LINER) from optical spectroscopy. This nearby
active galaxy (D=32 Mpc for $H_0 = 75 \rm \ Mpc^{-1} \ km \ s^{-1}$; 1\arcsec = 160 pc) is 
interesting as well for its nuclear starburst which is few Myr old as shown by its Wolf-Rayet
features
at 0.47 \micron\ (HeII).  This galaxy
has been studied in details in previous papers (Eckart et al. 1991,1996; Schinnerer
et al. 2000) at different wavelengths (NIR, X-ray, single dish CO).The study of NGC 6764 
has been motivated by the composite nature of its nucleus hosting a young powerful starburst 
as well as an AGN.

\section{Results}
The {\mbox{$^{12}$CO(1--0) }}and {\mbox{$^{12}$CO(2--1) }} lines have been observed with the IRAM 
interferometer 
located on the Plateau de Bure (France) with a spatial resolution of  $2.0\arcsec\times 
1.6\arcsec$ 
(PA=7$^\circ$) for the {\mbox{$^{12}$CO(1--0) }}data and $1.25\arcsec\times 1.25\arcsec$ 
for the {\mbox{$^{12}$CO(2--1) }} data. CO emission has been found in the very center of the 
galaxy 
($ R < 1.5$ kpc, about 10\arcsec) and the  total mass of molecular gas mass (including He) is 
found to be about 
$6.7\times 10^8 {\rm \ M_{\odot}}$.  
About 50 \% of the total molecular
gas mass is concentrated in one {\mbox{$^{12}$CO(1--0) }} beamsize in the center of NGC~6764. 
An offset in the position of 
the CO peak
and the 20cm radio continuum peak is observed ($\sim 1.2^{\prime\prime}$). The most striking
feature is the molecular gas flaring above the plane correlated with the 20cm radio continuum 
outflow starburst-drivem (see Fig. 1). The {\mbox{$^{12}$CO(1--0) }  line  is observed up to 
1 kpc  above the galactic plane. An
estimate gives a molecular gas mass of $1.7\times 10^7 {\rm \ M_{\odot}}$ out-of the plane, with 
z $>$ 500 pc ($\sim$1.5 CO(1-0)--beamsize).
The physical conditions of the molecular gas is probed by the 
CO(2-1)/CO(1-0) line ratio which is found to be about 1--1.3. Nevertheless a peak of 
about 2.0 for this line ratio  is found, spatially associated with the active nucleus, 
at  a blueshifted velocity of -140 km/s (rel. to the galaxy) when a lower CO line ratio is found
outside the central part (mainly the bar).
The nuclear CO kinematics show some molecular gas associated with bar orbits 
($x_2$ orbits).

\begin{figure}[t]
\centerline{
\psfig{figure=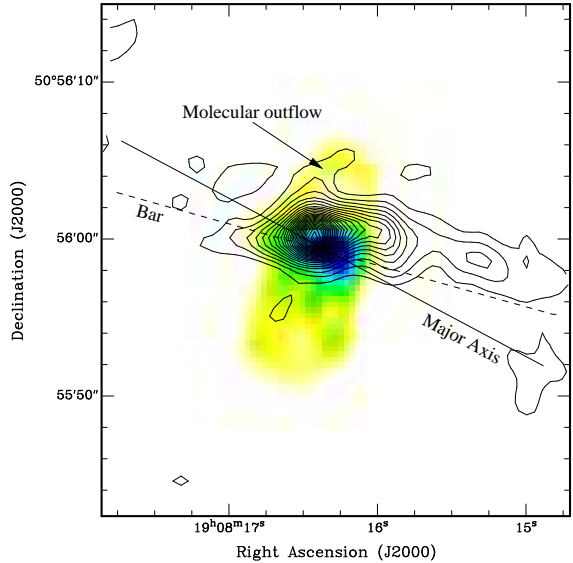,width=7.5cm}}
\caption{ VLA 20cm emission overlaid by $^{12}$CO(1--0) emission with contour levels from
1 to 27 {\mbox{Jy.beam$^{-1}$}}km/s  by step of 1 {\mbox{Jy.beam$^{-1}$}}km/s. The position 
of the major
axis and the bar are shown respectively by the solid and dashed lines.}
\label{fig1}
\end{figure}

\end{document}